# Guessing the upper bound free-energy difference between native-like structures

Jorge A. Vila[§]

[§]IMASL-CONICET, Universidad Nacional de San Luis, Ejército de Los Andes 950, 5700-San Luis, Argentina

Use of a combination of statistical thermodynamics and the Gershgorin theorem enable us to guess, in the thermodynamic limit, a plausible value for the upper bound free-energy difference between native-like structures of monomeric globular proteins. Support to our result in light of both the observed free-energy change between the native and denatured states and the microstability free-energy values obtained from the observed micro-unfolding tendency of nine globular proteins, will be here discussed.

**Introduction**

An accurate determination of the free-energy difference between native-like conformers of any protein is a daunting task. Indeed, studies of ubiquitin folding through *state of the art* equilibrium atomistic simulations[1] predict, at the melting temperature, a folding enthalpy (~14 kcal/mol) which is several times lower than the observed value (~84 kcal/mol).[2] Another case in point is the longstanding evidences[3] showing that the range of microstability free-energy values of native-like conformers of globular proteins is very narrow (2.5 to 7.1 kcal/mol), although, to be best of our knowledge, no theoretical proof supporting this assessment had been provided, yet. How to tackle the latter is illustrated here by analyzing the fluctuation around the native-state of monomeric globular proteins. For this purpose we make use of the following fact: the Gibbs free-energy of any protein state is given, in the thermodynamic limit, by the maximum eigenvalue of the partition function. Hereafter, an upper bound to the Gibbs free-energy difference between native-like states can be determined by using both the Gershgorin (circle) theorem[4] and a heuristic argument. Finally, the result is judged against both the observed free-energy change between the native and denatured states and the range of microstability free-energy values obtained from the observed micro-unfolding tendency of nine monomeric globular proteins.[5]

**Materials and Methods**

It is well known that *all* the thermodynamics properties of many biological problems of interest,[6] such as the helix-coil transition (induced by temperature or pH changes), loop entropy in RNA/DNA, etc., can be derived from the partition function ($Q$). Thus, in particular, the Gibbs free energy (G) will be given by:

$$G = -RT \ln Q \qquad (1)$$

After a proper assignment of statistical weight the partition function ($Q$) can be written in terms of a matrix ($\mathcal{C}$) where the elements are, indeed, Boltzmann factors.[6] Then, it is well known[7] that, in the thermodynamic limit, the following equality hold:

$$\mathcal{L}im_{j \to \infty} (1/j) \ln Q = \mathcal{L}im_{j \to \infty} (1/j) \ln C_{kl}(j) = \ln \lambda_{max} \qquad (2)$$

where $C_{kl}(j)$ is any element of the iterate matrix $\boldsymbol{C}^j$; $j$ is the number of residues in the chain and $\lambda_{max}$ is the maximum eigenvalue of the matrix $\boldsymbol{C}$.

According to the Gershgorin (circle) theorem,[4] for any eigenvalue $\lambda$ of the matrix $\boldsymbol{C}$ the following inequality hold $|\lambda| \leq \max\{\sum_{l=1}^{n}|C_{kl}|\}$ with $1 \leq k \leq n$ and $n$ the matrix order. Taking into account that the maximum eigenvalue of $\boldsymbol{C}$ is real and positive and that any element $\boldsymbol{C}$, by definition, satisfies $|C_{kl}| \equiv C_{kl}$, then there is a $k'$ value such that the upper bound of the maximum eigenvalue ($\lambda_{max}$) is given by:

$$\lambda_{max} \leq \{\sum_l C_{k'l}\} \qquad (3)$$

Assuming that the native-like conformers of a given ensemble coexist in fast dynamics equilibrium,[8] then an upper bound to the free-energy difference, between conformers with the lowest and highest total free-energy $G$, can be computed from equations (1) to (3), as:

$$\Delta G \leq \mathcal{Lim}_{j \to \infty} RT \ln \left[\sum_l C_{k'l} / \sum_m C_{t'm}\right]^j \qquad (4)$$

where the term $[\sum_l C_{k'l} / \sum_m C_{t'm}] > 1$ because the lowest free energy conformation will be, according with the thermodynamic hypothesis,[9] the native-state. Implicit in this inequality is that *all* conformers, in equilibrium with the native-state, posses comparable total free-energy. Indeed, if equation (4) were use to guess an upper bound to the free-energy difference between *native* and *non-native* (denaturated) states, then it is reasonable to assume that $[\sum_l C_{k'l} / \sum_m C_{t'm}] \gg 1$. However, analysis of the protein unfolding is out of the scope of this work.

**Results and Discussion**

There are two dominant interactions that contribute to the stability of native-like conformers in proteins, regardless the fold class, sequence or size, specifically, interactions between *i*) polar grups (hydrogen bonds) and *ii*) non-polar groups.[10-13] Consequently, the free-energy changes between native-like conformers, given by equation (4), would imply variations of either one, or both, interactions. While we recognize there could be many functional forms involving this two interactions, it called our attention that the molecular weight (*MW*), of nine monomeric globular proteins,[5] show a good correlation with the total number of both the intramolecular hydrogen bonds ($R^2 = 0.98$) and the pairs of nonpolar groups at distances < 4Å ($R^2 = 0.83$). Thus, from a heuristic point of view, we conjecture that the term $[\Sigma_l\, C_{k'l} / \Sigma_m\, C_{t'm}]^j$ grows with *j* as molecular weight (*MW*) does. Then, we can rewrite equation (4) as:

$$\Delta G \leq \mathcal{L}im_{MW \to \infty} RT \ln MW \qquad (5)$$

Considering that the largest known monomeric globular protein[14] possess a $MW = 2.7\ 10^5$, then, at T = 298K, the equation (5) give us a $\Delta G \leq 7.4$ kcal/mol. This plausible value for ΔG, which is robust upon small *MW* changes, represents the upper bound for the free-energy difference between native-like structures of monomeric globular proteins.

At this point it is worth noting that the observed free energy values of microstability (micro-unfolding) determined from nine monomeric globular proteins[5] satisfy the inequality $\Delta G \leq 7.1$ kcal/mol and that the corresponding average free-energy of denaturation (macro-unfolding) is $<\Delta G> \cong 11 \pm 3$ Kcal/mol. The plaussible value for the upper bound free-energy difference ($\Delta G \leq 7.4$ kcal/mol) is, certainly, in line with these results.

**Conclusions**

In summary, based on the use of simple concepts of statistical thermodynamics and the Gershgorin theorem, that offer an straightforward way to estimate the upper bound for the maximum eigenvalue of a matrix, we have been able to predict, by using a heuristic argument, a plausible value for the largest free energy difference between coexistent native-like structures of monomeric globular proteins, namely 7.4 Kcal/mol. Support to this result is provided by the demonstrated consistency with the observed free-energy changes from a set of nine globular proteins.

Considerable attention has been dedicated during the last 40 years to develop methods with which to compute the free energy of biological systems accurately. In this regards, the work proposed herein may spur significant progress for the development of new methods for free energy calculations aimed at solving problems of paramount importance such as an unambiguous characterization of the protein folding, misfolding and aggregation.

**Acknowledgments**

This research was supported by CONICET-, the UNSL- and ANPCyT-Argentina.

**Author Contributions**

J.A.V designed, conceptualized the project and wrote the manuscript.

**Competing Financial Interest**

The author declares no competing interest.